\documentclass[letterpaper,aps,prc,superscriptaddress,nofootinbib,showpacs,floatfix,reprint]{revtex4-1}
\usepackage[utf8]{inputenc}
\usepackage{amsmath}
\usepackage{amsfonts}
\usepackage{amssymb}
\usepackage{amsthm}
\usepackage{bm}
\usepackage[english]{babel}
\usepackage{fontenc}
\usepackage{graphicx}
\usepackage[margin=1in]{geometry}
\usepackage{enumerate}
\usepackage{color}

\newcommand{\Lc}{${\Lambda_{\rm c}^+\;}$}
\newcommand{\D}{$\text{D}^0\;$}

\newcommand{\pt}{$p_{\rm T}\;$}
\newcommand{\ee}{$\rm e^-e^+$}
\newcommand{\ep}{$\rm e^-p\;$}

\newcommand{\snn}{$\sqrt{s_{\text{NN}}} = 13 \; \text{TeV}$}

\definecolor{darkgreen}{rgb}{0.0, 0.6, 0.22}
\definecolor{dblue}{rgb}{0.,0.,1.}       
\definecolor{orange}{cmyk}{0.,0.453,1.,0.}    

\begin{document}
\title{Probing Partonic Evolution and Hadronization via Balance Functions and Correlations of Charmed Hadrons}

\author{ Oveis~Sheibani }
\email{oveis.sheibani@cern.ch}
\author{ Claude~Pruneau }
\email{claude.pruneau@wayne.edu}
\author{ Victor~Gonzalez }
\email{victor.gonzalez@cern.ch}
\affiliation{Department of Physics and Astronomy, Wayne State University, Detroit, 48201, USA}
\author{ Sumit~Basu }
\email{deceased}
\affiliation{Department of Physics, Lovely Professional University, Phagwara, India}
\author{Alexandru Florin Dobrin}
\affiliation{Institute of Space Science - INFLPR Subsidiary, Magurele, 077125, Romania } 
\email{alexandru.florin.dobrin@cern.ch}
\author{Yash Patley}
\email{yashpatley.iitb@gmail.com}
\author{Basanta Nandi }
\email{basanta@phy.iitb.ac.in}
\author{ Sadhana Dash }
\email{sadhana@phy.iitb.ac.in}
\affiliation{Department of Physics, Indian Institute of Technology
 Bombay, Mumbai - 400076, India }

\date{\today}

\begin{abstract}
Predictions of charm correlation functions and more specifically balance functions are presented in proton--proton (pp) collisions at \snn\ based on the \textsc{Pythia} 8.3 event generator. Correlations are  computed for  identical and  cross species charmed hadrons in both minimum bias and high-\pt biased collisions. We study the strength of correlations as a function of the number of balanced flavors and  investigate the impact of variations of \textsc{Pythia} parameters controlling the Lund string fragmentation on the shape and strength of the correlation functions. The feasibility of measurements of the charm balance function presented is discussed in the context of the future LHC  experiments.
\end{abstract}
\maketitle

\section{Introduction}
\label{sec:introduction}
Measurements of heavy-flavor production conducted in recent years  in  high-energy pp collisions suggest the possibility of a breakdown of the factorization approach used to compute particle production cross sections. It emerges that particle production and transport cannot be simply scaled from \ee and \ep collisions. Whether this breakdown can be ascribed to non-perturbative corrections beyond standard fragmentation functions or manifestations of collective behavior similar to that observed in Pb--Pb collisions is however a subject of ongoing investigation. Accordingly, various pieces of evidence, including inclusive cross sections, elliptic flow, and particle correlations, might contribute to shaping the  understanding of this phenomenon.  

Correlations involving heavy-flavor hadrons are particularly valuable because they retain information about the early stages of the collisions and to a certain extent preserve this information during transport and hadronization. Given their relatively large masses, charm and bottom quarks yields are not significantly affected by subsequent thermal production. The question is thus whether observed correlations of charm hadrons with other particles are shaped by their production mechanism or by transport through a medium in pp collisions as well as in A--A collisions. Naively, observed  separations in rapidity  suggest that production plays a significant role. However, transport effects might have a noticeable impact on final-state hadron correlations. 

Predictions of  charmed hadron production are typically based on the factorization hypothesis which allows to use tuned fragmentation functions obtained from \ee  and \ep data to extrapolate calculations  to the  more complex processes involved in pp and A--A collisions.  However, recent data from small-system collisions (pp and p–Pb) at LHC energies suggest a potential breakdown of the factorization ~\cite{ALICE:2021dhb}. This is suggested by the observation of a  deviation  in the $p_{\rm T}$-integrated ratios of open-charm hadron species where \D yield is suppressed relative to \ee collisions, whereas baryon production, particularly \Lc and other charmed baryons such as $\Xi_{c}^{0}$ and $\Xi_{c}^{+}$, is enhanced. Conversely, nucleon--nucleon collisions at high energies favor the production of less abundant charmed hadron than basic \D  compared to observations in \ee  and \ep collisions. The $p_{\rm T}$-differential baryon to meson ratio significantly deviates from the \ee trend, exhibiting a multiplicity dependence of 5.3$\sigma$ in pp collisions~\cite{ALICE:2021npz}, the ratios were also measured as function of multiplicity in p-Pb~\cite {ALICE:2022exq, ALICE:2020wla, Sheibani:2023qed}.  These findings suggest that the nucleon--nucleon medium, especially at high multiplicities, is denser and richer than previously thought.

Measurements in central Pb--Pb collisions revealed substantial suppression of charmed hadrons, quantified by the nuclear modification factor $R_{AA}$, with prompt D-meson yields reduced by up to a factor of 5 at $p_{\rm T} = 6$--$8$~GeV/$c$, and D$_s^+$ mesons showing comparable suppression at higher \pt{}, signaling charm quark energy loss in the quark--gluon plasma (QGP)~\cite{ALICE:2015vxz} Moreover, finite elliptic flow coefficients ($v_2$) of D mesons and $\mathrm{J/\psi}$~\cite{ALICE:2020iug, ALICE:2023gjj, ALICE:2020pvw, CMS:2018duw} indicate that charm quarks participate in the collective expansion of the medium. Charmonium states such as $\mathrm{J/\psi}$ can either be suppressed at high temperatures or regenerate through recombination at lower temperatures, with sequential suppression patterns at RHIC and LHC supporting these regeneration effects~\cite{STAR:2019fge, Feuillard:2023xgh, CMS:2018zza}. Results from p--Pb collisions show intermediate suppression and flow behavior, suggesting partial medium-like effects in smaller collision systems as well. These findings imply that, despite their early production, the final-state distributions of charm quarks are sensitive to the surrounding medium. This raises the question of whether high-multiplicity pp collisions may also exhibit such medium-induced effects. However, flow measurements in small collision systems are hindered by the difficulty of defining symmetry planes and the short-lived nature of the medium.

The notion of balance function (BFs) was introduced in Ref~\cite{Pratt:2011bc}, but subsequently modified into general and unified BFs that enable studies of system with non-vanishing net charge and ab initio removal of non correlated pair yields \cite{Pruneau:2022mui, Pruneau:2023cea}. Balance functions feature an explicit dependence on conservation laws, including conservation of energy, momentum, electric charge, baryon number, and flavor conservation. This explicit dependence on conservation laws constrains the strengths and shapes of balance functions making them sensitive to the time evolution of particle production, transport, regeneration, and hadronization mechanisms. Charm--anticharm BFs and $\mathrm{J/\psi}$--charm correlation functions can probe early gluon-dominated stages of heavy-ion collisions. These observables are sensitive to initial-state configurations as well as the collision dynamics. They thus offer experimental access to the hypothesized gluon plasma phase before chemical equilibration occurs~\cite{Basu:2021dzv}.

The goal of this work is to explore whether charm balance functions can provide further insights into the production and transport of charm quarks throughout the pp collision life cycle and whether they are measurable in the context of ongoing experiments or must await next generation technologies. Section~\ref{sec:motivation} elaborates the motivation for this work while Sec.~\ref{sec:analysismethods} describes the simulations and methods used to obtain charm-hadron correlation functions. Results of simulations of charm correlations based on normalized differential cumulants and BFs are presented and discussed in Sec.~\ref{sec:results}. A summary and conclusions of this work are presented in Sec.~\ref{sec:conclusion}.

\section{Motivation and Scope}
\label{sec:motivation}
In the Lund string fragmentation model, the charm quark hadronizes by combining with a light quark and antiquark produced via string breaking. These light quarks effectively form a closed color flux tube with the charm quark. Their correlation may depend on their proximity in phase space—e.g., being kinematically close or within the same shower—or they may arise from separate, independent sources in the event; in such cases, the parton closing the color flux tube with the charm quark could emerge through long-range non-perturbative interactions, which is modeled in \textsc{Pythia} via color reconnection. The origin of the light quark influences both the flavor content and the kinematics of its associated charm quark which will be manifested in their correlation. Light quarks produced via gluon splitting in association with the $c\bar{c}$ pair tend to be kinematically close to the charm quark companion and follow a flavor hierarchy governed by quark mass and available phase space, thereby resulting in a strong correlation. In contrast, light quarks from independent or medium-induced sources are more randomly distributed in both flavor and momentum space, and should thus lead to weaker or longer reach correlations. Examining how the correlation strength varies among different balancing flavor pairs~\cite{Patley:2024egb} should thus provide information about the phases of the dynamics of the heavy quarks and of the light quarks that get confined together. Localized processes like gluon splitting are expected to yield narrower correlations in phase space, while broader distributions would be indicative of enhanced flavor mixing or spatial decorrelation. In this study, we employ the \textsc{Pythia}8 model to compute charm correlation and balance functions in order to  explore whether these could provide new insights into charm production and evolution.
 
Theoretical studies of BFs require models that properly integrate  conservation laws. While this is done ab initio in string breaking models such as \textsc{Pythia}~\cite{Bierlich:2022pfr} and HERWIG~\cite{Gieseke:2003hm}, or in transport models such as UrQMD~\cite{Petersen:2008dd}, it is more difficult to achieve in models describing the evolution of a potential medium based on  hydrodynamics with final state hadron production by means of a Cooper-Frye particlization ansatz~\cite{Cooper:1974mv}. Cooper-Frye particlization, though sufficient for calculations of anisotropic flow, introduces event-averaging that locally violates quantum number conservation and is thus not suitable for the computation of trustable BFs. An additional complication arises in the context of hydrodynamic frameworks because the production of charmed hadrons cannot be treated thermally. Heavy-flavor production is strongly suppressed by canonical effects due to the large charm mass and the low thermal charm yield at typical medium temperatures. New techniques, such as those used in EPOS-HQ, are thus required to model heavy-flavor production in this context~\cite{Gossiaux:2019syn}. Additionally, although  considerable advances have been accomplished towards particlization schemes that locally or semi-locally conserve all quantum numbers, further work is needed to assess whether these new  schemes can properly account for long range correlations, particularly those of baryons~\cite{Oliinychenko:2020cmr}. From a  practical standpoint, it must also be acknowledged that the simulation of A--A collisions with $3+1$ hydrodynamic models required to achieve a detailed description of charge inclusive and charge dependent correlation functions is extremely CPU intensive, whereas the relative simplicity and efficiency of the \textsc{Pythia} model enables a reasonably rapid generation of the vast number of simulated events required for detailed differential studies of hadron--hadron correlation functions, and more specifically charm BF. The study of charm BF presented in this work is thus limited to the \textsc{Pythia}8 model. We will in fact show that the realization of calculation of charm BF nonetheless require considerable computing resources. Given the production of pp collisions with \textsc{Pythia} can be biased towards mechanisms involving higher momentum transfers, we also study the impact of such bias on the strength and shape of BF.

\section{Analysis Methods}
\label{sec:analysismethods}
Charm correlations studies reported in this work are based on  differential cumulant correlation functions and more specifically   general balance functions~\cite{Pruneau:2022mui}. They are expressed in terms of single and pair densities denoted respectively $\rho_1^{\alpha}(\vec p)$ and $\rho_2^{\alpha\beta}(\vec p_1, \vec p_2)$, where $\alpha$ and $\beta$ indicate the particle species of interest. The densities are integrated over the transverse momenta of particles and evaluated as event ensemble average of single and pair yields computed  as functions of particle pair longitudinal $\Delta y$ and azimuthal $\Delta \varphi$ separations. The two-particle correlation function is defined as: 
\begin{widetext}
\begin{align}
\label{eq:C2}
C_2^{\alpha\beta}(\Delta y, \Delta \varphi) = \rho_2^{\alpha\beta}(\Delta y, \Delta \varphi) - 
\rho_1^\alpha\otimes \rho_1^\beta(\Delta y, \Delta \varphi).
\end{align}
\end{widetext}
The strength of correlations can also be characterized using the normalized two-particle differential cumulant $R_2^{\alpha\beta}(\Delta y, \Delta \varphi)$, defined as:
\begin{align}
\label{eq:R2}
R_2^{\alpha\beta}(\Delta y, \Delta \varphi) =
\frac{C_2^{\alpha\beta}(\Delta y, \Delta \varphi)}
{\rho_1^{\alpha}\otimes \rho_1^{\beta}(\Delta y, \Delta \varphi)} - 1,
\end{align}
where the normalization is performed with respect to the product of single-particle densities. The functions $C_2^{\alpha\beta}(\Delta y, \Delta \varphi)$ and $R_2^{\alpha\beta}(\Delta y, \Delta \varphi)$ vanish identically in the absence of correlations. The analysis can then be entirely focused on truly correlated particles. Such correlations can arise from energy-momentum, charge, baryon number, and flavor conservation, as well as transport processes. In A--A collisions, these might include longitudinal, radial flow, as well as anisotropic flow. Such transport phenomena are, however, not expected from a model like \textsc{Pythia}. 

Balance functions are computed in terms of {\bf associated particle densities}, also known as {\bf conditional cumulants}, $A_2^{\alpha|\beta}(\Delta y, \Delta \varphi)$, defined according to  
\begin{align}
\label{eq:A2}
A_2^{\alpha|\beta}(\Delta y, \Delta \varphi) = 
\frac{C_2^{\alpha\beta}(\Delta y, \Delta \varphi)} 
{\langle N_1^{\beta} \rangle},
\end{align}
in which $\beta$ represents the {\bf reference} particle species and $\alpha$ its {\bf associate}~\footnote{The notation $\alpha|\beta$ is to be interpreted as measuring a particle of species $\alpha$ given a particle of species $\beta$ is observed}. The quantity $A_2^{\alpha|\beta}$ reflects the associated yield per reference particle, whereas $R_2^{\alpha\beta}$ measures the strength of correlations relative to single particle yields. $R_{2}$  is experimentally of interest because it is relatively insensitive to trivial instrumental effects like acceptance and efficiency. BFs, however, are of specific interest because they must satisfy sum rules determined by quantum number conservation. They are defined   according to 
\begin{widetext}
\begin{align}
B_2^{\alpha|\bar\beta}(\Delta y, \Delta \varphi)
= A_2^{\alpha| \bar{\beta}}(\Delta y, \Delta \varphi)
-A_2^{\bar\alpha|\bar\beta}(\Delta y, \Delta \varphi), 
\end{align}
\end{widetext}
in which barred label $\bar\alpha$, $\bar\beta$ identify the antiparticles of species $\alpha$ and $\beta$, respectively. The underlying assumption is that correlated pairs $\bar\alpha,\bar\beta$ and $\alpha,\bar\beta$ are similarly affected by energy-momentum conservation and all other aspects of particle production and differ primarily (or only) on the basis of the quark pair flavor creation process. BFs thus enable the study of the charge (or flavor, or baryon) balancing process: given a species $\bar\beta$ of charge $-1$ is observed at $\vec p_2$, where is, on average, its balancing partner of charge $+1$.  By construction, general balance functions satisfy simple rules when integrated over the full phase space of particle production~\cite{Pruneau:2022mui}.  General balance functions have sensitivity to the particle production processes, collective transport (e.g., flow) as well as rescattering and the diffusivity of light quarks. The relative strength of integrals of balance functions considered over a finite acceptance is additionally sensitive to the hadro-chemistry of the particle production~\cite{Pruneau:2019baa}. In this work, correlation $C_2$, $A_2$, and balance functions are integrated over the full range of transverse momentum production, $p_{\rm T}>0$, the full azimuth, $0\le \varphi < 2\pi$, as well as the full range of particle production in rapidity. 
 
\begin{figure}[htbp]
  \centering
  \vspace{2pt}
  \includegraphics[width=\columnwidth,clip]{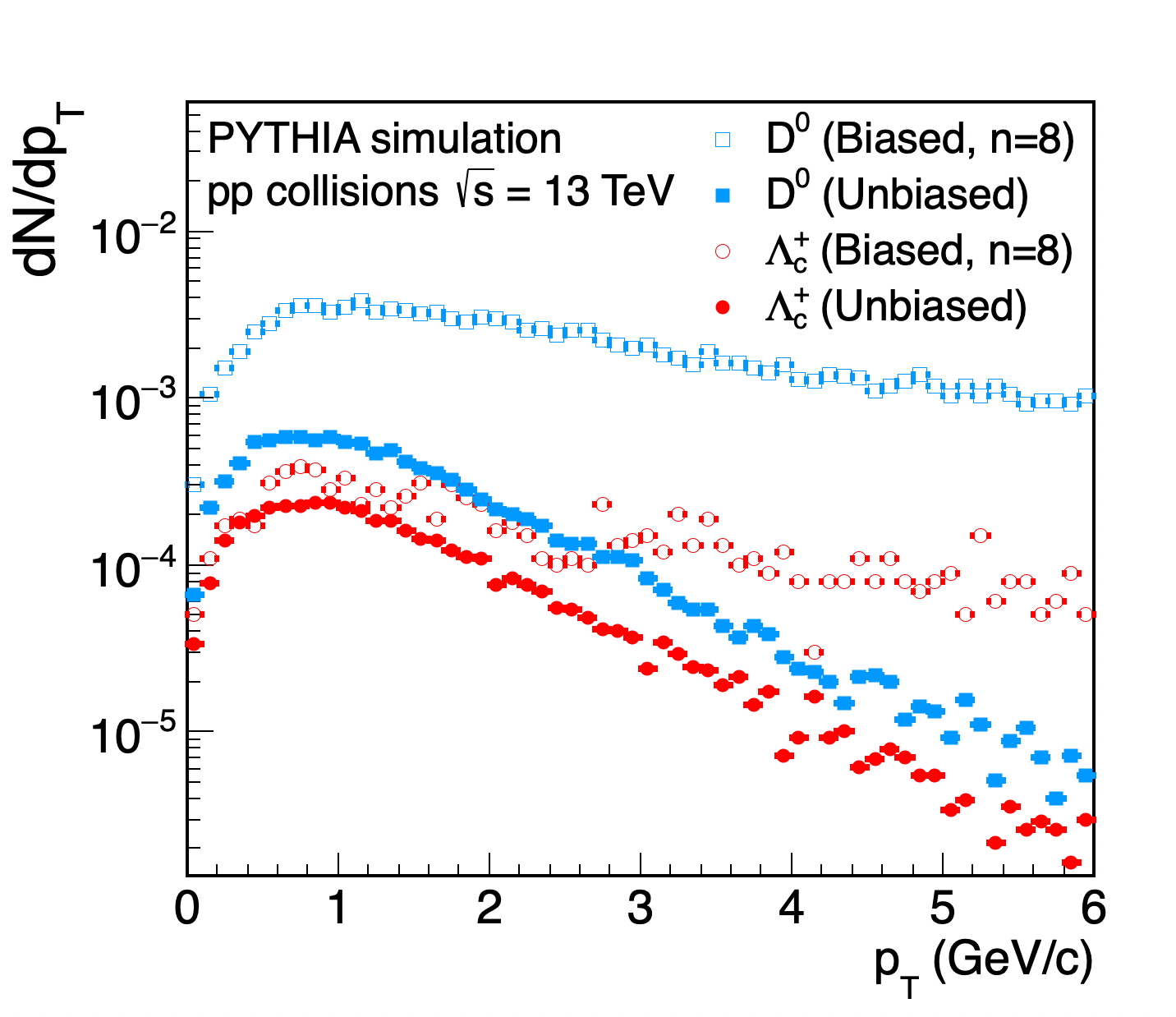}
\caption{Transverse momentum distributions ${\rm d}N/{\rm d} p_{\rm T}$ of $\mathrm{D^0}$ and $\mathrm{\Lambda_c^+}$ hadrons obtained with momentum scale reweighed (open symbols)  relative to unbiased events (solid symbols) in pp collisions at $\sqrt{s} = 13$~TeV simulated with  \textsc{Pythia} 8.3. The distributions are normalized to the number of events. }
\label{Single}
\end{figure}

Simulations carried out in this work are based on the \textsc{Pythia} 8.3 generator which, by construction, guarantees explicit conservation of all quantum numbers as well as energy and momentum. Calculations of charm balance are thus meaningful because the creation of a charm quark is explicitly balanced by the creation of anticharm quark locally and on an event-by-event basis. However, the production of statistically meaningful unbiased charm hadrons correlations is a rather computing intensive task given the small charm production cross section in pp collisions. Simulations were conducted on the Wayne State computing grid which enabled reasonably efficient production of a data set of 50 billion \textsc{Pythia} events. As we show below, the statistical significance of the charm--charm and charm--anticharm correlation functions remains somewhat poor even with such a large data sample. Given the charm production increases with the effective momentum scale, we have thus sought to improve the statistical accuracy of the correlation functions  by introducing a momentum scale bias based on a mechanism native to the \textsc{Pythia} model which enables a reweighing of the cross section according to 
\begin{equation}
\label{eq:biasing}
    \frac{{\rm d} \sigma_{\text{Biased}}}{{\rm d} \sigma_{\text{Unbiased}}} = \left(\frac{\hat{p}_{\mathrm{T}}}{\hat{p}_{\mathrm{T},\text{Ref}}}\right)^n,
\end{equation}
where $\hat{p}_{\mathrm{T}}$ is the momentum scale of particle production processes. Simulations were carried out with  $\hat{p}_{\mathrm{T}}\sim 10$ GeV/$c$ and with values $n=2$ and $n=8$. With the momentum reweighing, it was possible to obtain statistical accuracy for the computation of charm--charm correlation functions based on 5 billion and 
200 million events. The use of this bias not only improved the computational efficiency but also enabled studies of correlation functions for higher momentum scale processes. Separating soft and hard processes is essential for exploring different underlying physics mechanisms, as charm production and hadronization may proceed differently in soft-dominated versus hard-scattering events. The impact of the reweighing is illustrated in Fig.~\ref{Single} which compares transverse momenta spectra of $\mathrm{D^0}$ and $\mathrm{\Lambda_c^+}$, obtained with the unbiased and biased ($n=8$) samples obtain in pp collisions at $\sqrt{s} = 13$~TeV. One finds indeed that the reweighing procedure enhances the charm production per event by more than one order of magnitude relative to the unbiased sample. 

Particle production in \textsc{Pythia} proceeds through hard QCD processes, tuned multiparton interactions, showers, color reconnection mechanisms, and hadronization. Of particular interest in this analysis is the production of charm via hard processes and the subsequent formation of Lund strings~\cite{Andersson:2001yu}. In this context, it is especially intriguing to investigate whether the longitudinal width of correlations are sensitive to the parameters governing string formation, particularly the Lund fragmentation parameters \(a\)-Lund and \(b\)-Lund. To address this, we generated \textsc{Pythia} events using several configurations with varied \(a\) and \(b\) values, in order to explore how changes in these parameters may be reflected in the properties of charm correlations. Table~\ref{tab:parameters} presents the main \textsc{Pythia} parameters adjusted in this analysis, along with their values and the physics processes they affect. Parameters such as $a$-Lund and $b$-Lund control the Lund fragmentation behavior, while color reconnection influences final-state particle configurations. 
\begin{table}[tp]
\caption{Lund string parameters considered in this work.}
\begin{tabular}{|l|c|c|}
\hline
\textbf{Parameter} & \textbf{Value} & \textbf{Physics} \\ \hline
\texttt{expPow} & 1.85 & Multiparton interactions \\ \hline
\texttt{pT0Ref} & 2.15 & Multiparton interactions \\ \hline
\texttt{a-Lund} & 0.36 & Lund string fragmentation \\ \hline
\texttt{b-Lund} & 0.56 & Lund string fragmentation \\ \hline
\texttt{probQQtoQ} & 0.078 & Quark flavor production \\ \hline
\texttt{ProbStoUD} & 0.2 & Strange-to-up/down suppression \\ \hline
\texttt{Reconnection} & mode 2& Color reconnection \\ \hline
\end{tabular}
\label{tab:parameters}
\end{table}

The correlation functions $A_2$ and $B_2$ are defined as combinations of two or more terms that may be statistically correlated in the Monte Carlo generation process. Rather than relying on standard error propagation techniques, which can be inadequate in the presence of such correlations, statistical uncertainties were estimated using the sub-sample technique. In this method, the full dataset is divided into several statistically independent sub-samples, the correlation functions are computed separately within each sub-sample, and the resulting spread is used to calculate the final uncertainties as the standard deviation across sub-samples. 

\section{Results and discussion}
\label{sec:results}

\begin{figure*}[htbp]
  \centering
  \vspace{2pt} 
   \includegraphics[width=\linewidth]{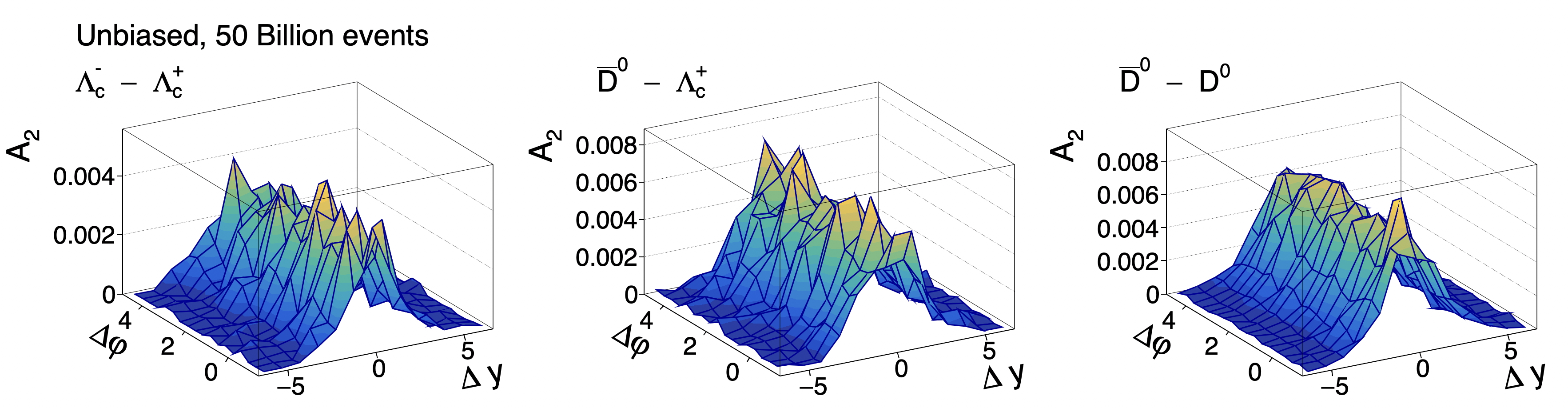}
\caption{Two-particle associated cumulants $A_2$ obtained for hadron pairs $\mathrm{\Lambda_c^-\Lambda_c^+}$ (left), $\mathrm{\overline{D}^0\Lambda_c^+}$ (middle), $\mathrm{\overline{D}^0D^0}$ (right) with no momentum transfer selection.}
\label{fig:2D0}
\end{figure*}

\begin{figure*}[htbp]
  \centering
  \vspace{2pt}  
  \includegraphics[width=\linewidth]{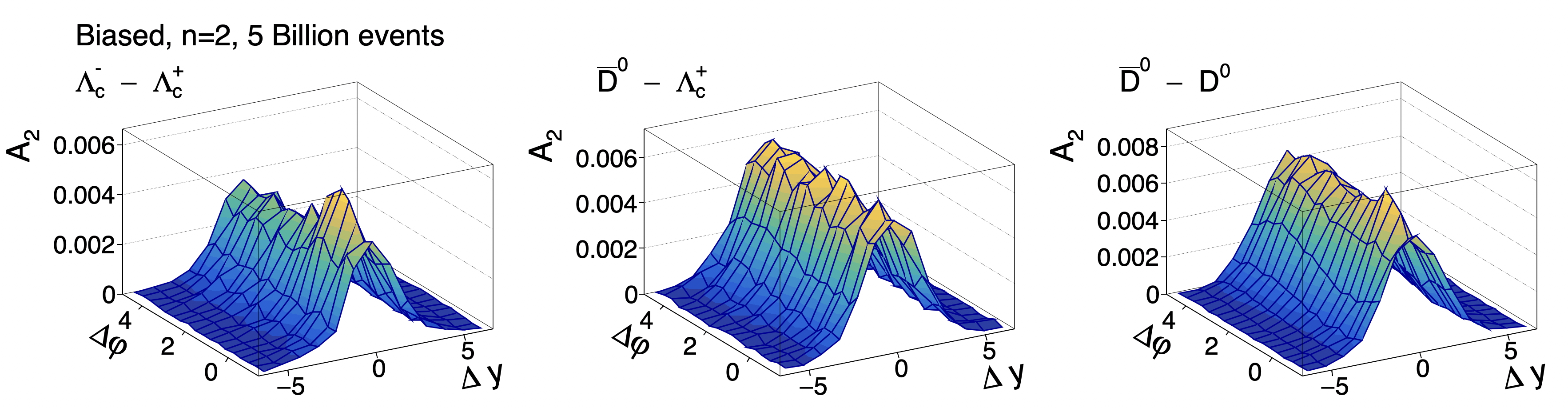}
\caption{Two-particle associated cumulants $A_2$ obtained for hadron pairs $\mathrm{\Lambda_c^-\Lambda_c^+}$ (left), $\mathrm{\overline{D}^0\Lambda_c^+}$ (middle), $\mathrm{\overline{D}^0D^0}$ (right) with momentum transfer selection bias based on Eq.~(\ref{eq:biasing}) with $n=2$. }
\label{fig:2D1}
\end{figure*}

\begin{figure*}[htbp]
  \centering
  \vspace{2pt}  
  \includegraphics[width=\linewidth]{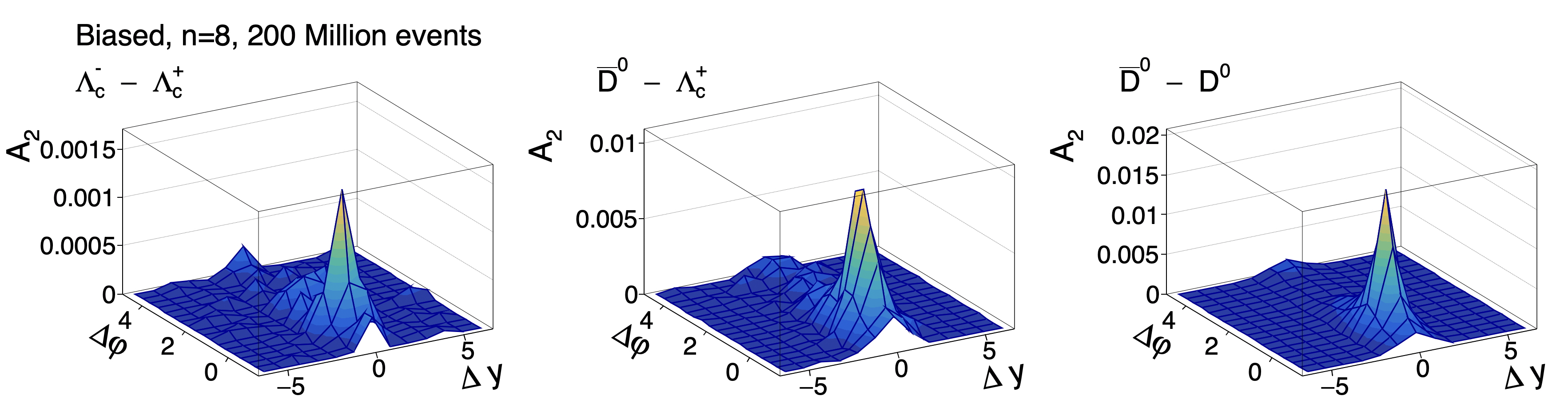}
\caption{Two-particle associated cumulants $A_2$ obtained for hadron pairs $\mathrm{\Lambda_c^-\Lambda_c^+}$ (left), $\mathrm{\overline{D}^0\Lambda_c^+}$ (middle), $\mathrm{\overline{D}^0D^0}$ (right) with momentum transfer selection bias based on Eq.~(\ref{eq:biasing}) with $n=8$.}
\label{fig:2D2}
\end{figure*}

Figures~\ref{fig:2D0}--\ref{fig:2D2} present correlation functions $A_2$ plotted as a function of the relative rapidity ($\Delta y$) and azimuthal angle difference ($\Delta \varphi$), for selected pairs of charmed hadrons in $\mathrm{pp}$ collisions at $\sqrt{s} = 13$ TeV. Figure~\ref{fig:2D0} shows correlations obtained for events generated with no momentum transfer scale bias, whereas Figs.~\ref{fig:2D1} and \ref{fig:2D2} represent mildly biased events obtained with $n = 2$ and highly biased events generated with $n=8$, respectively, both obtained with $\hat{p}_{\mathrm{T}}=10$ GeV/$c$. The azimuthal dependence between charm and anticharm hadron pairs is approximately uniform under minimum-bias conditions, indicating no preponderance towards  near-side or away-side pair emission. In sharp contrast, correlation functions $A_2$ shown in Figs.~\ref{fig:2D1} and \ref{fig:2D2} exhibit a strong near side peak.

In \textsc{Pythia}, $c\bar{c}$ pairs originate from a mixture of leading-order gluon fusion and other processes such as gluon splitting or flavor excitation. Gluon fusion is additive in nature: two incoming gluons produce a virtual gluon in a $2 \rightarrow 2$ hard scattering, which decays into a $c\bar{c}$ pair. Gluon splitting, by contrast, is fragmentative: a high-energy gluon radiated from a parton  subsequently splits into a $c\bar{c}$ pair. At different values of $\hat{p}_{\mathrm{T}}$, both mechanisms contribute, but with distinct topologies. \textsc{Pythia} includes all these mechanisms through a combination of matrix elements and parton showers, whose relative contributions shape the final-state topology. Parida et al.~\cite{Parida:2023qju} further demonstrated that, in pp collisions, final-state $c\bar{c}$ azimuthal correlations remain nearly flat, while in heavy-ion collisions, medium effects narrow $\Delta\varphi$ due to collective flow and broaden $\Delta y$ due to diffusion. The figures show that the azimuthal correlation between charm and anticharm hadrons evolves from a uniform distribution in unbiased events to a distinct near-side peak as the event bias increases, particularly visible in Fig.~\ref{fig:2D2} where the bias is the strongest. This reflects an increase in the momentum transfer of the hard scattering, where the outgoing gluons carry greater transverse momentum, producing $c\bar{c}$ pairs that are more collimated. As a result, the resulting hadrons such as $\mathrm{\Lambda_c^{\pm}}$ and $\mathrm{D}$ mesons inherit the focused geometry of their parent gluon, leading to a pronounced near-side enhancement in the final-state correlation.

One finds that the cumulants $A_2$, shown in Figs.~\ref{fig:2D0}--\ref{fig:2D2}, feature appreciable  statistical fluctuations in spite of the large pp collision sample generated and analyzed. That said, one notes, even with large correlation yield fluctuations  that the (unbiased) $A_2$ correlation functions of the species pairs considered have no obvious dependence on the relative angle of emission of the particles in $\Delta \varphi$. In the longitudinal direction, one notes the balancing pair correlations are strongest for small values of the rapidity pair separation $\Delta y$. The correlations are however somewhat broad and feature rms values of the order of 0.5-0.9. The rms width of the $A_2$ correlations is found to depend on the pair partners and the species of the reference hadron.

\begin{figure*}[hpt]
    \centering
    \vspace{2pt}
    \includegraphics[width=\linewidth,clip]{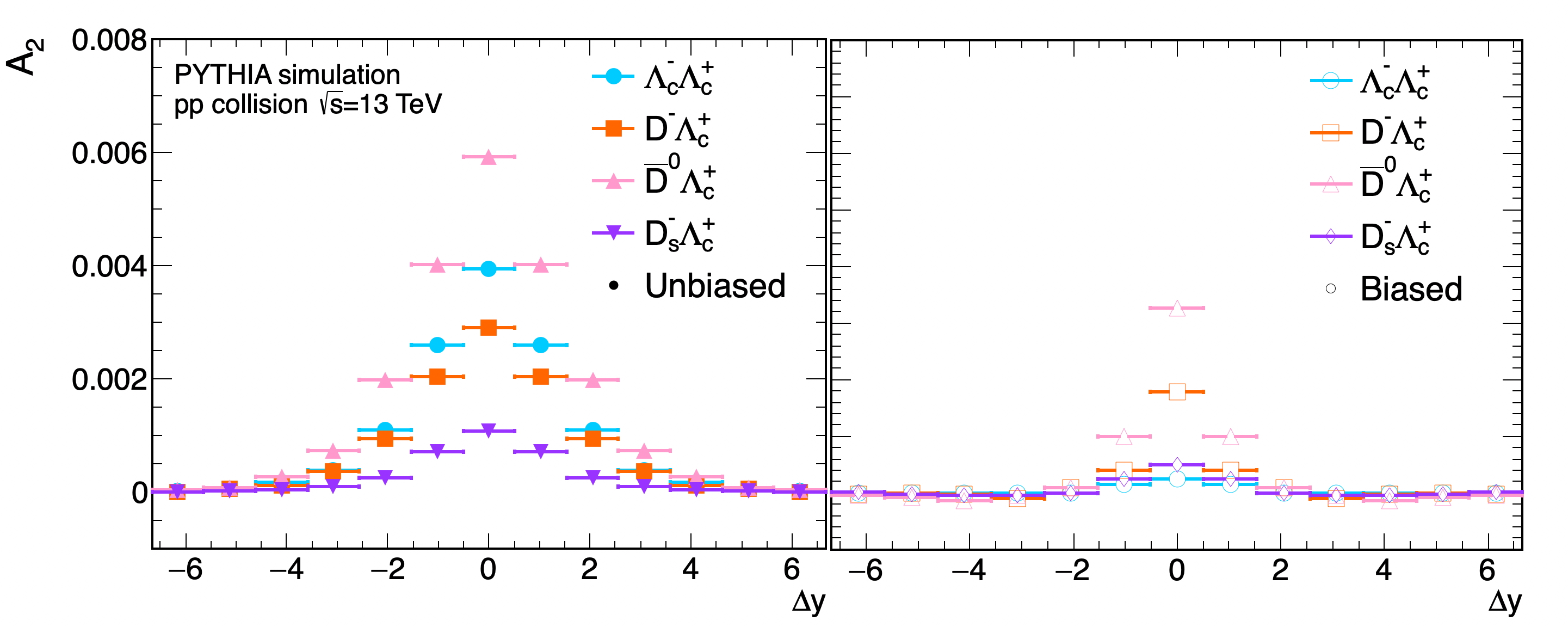}
\caption{Projections of the associated cumulants $A_2$ onto the $\Delta y$-axis  for selected charm--charm hadron pairs in \textsc{Pythia} unbiased (left) and biased generated based on Eq.~(\ref{eq:biasing}) with $n=8$ (right) events.}
\label{A2PorjectionUnbiasedandBiased}
\end{figure*}

\begin{figure}[hpt]
    \centering
    \vspace{2pt}
    \includegraphics[width=0.95\columnwidth,clip]{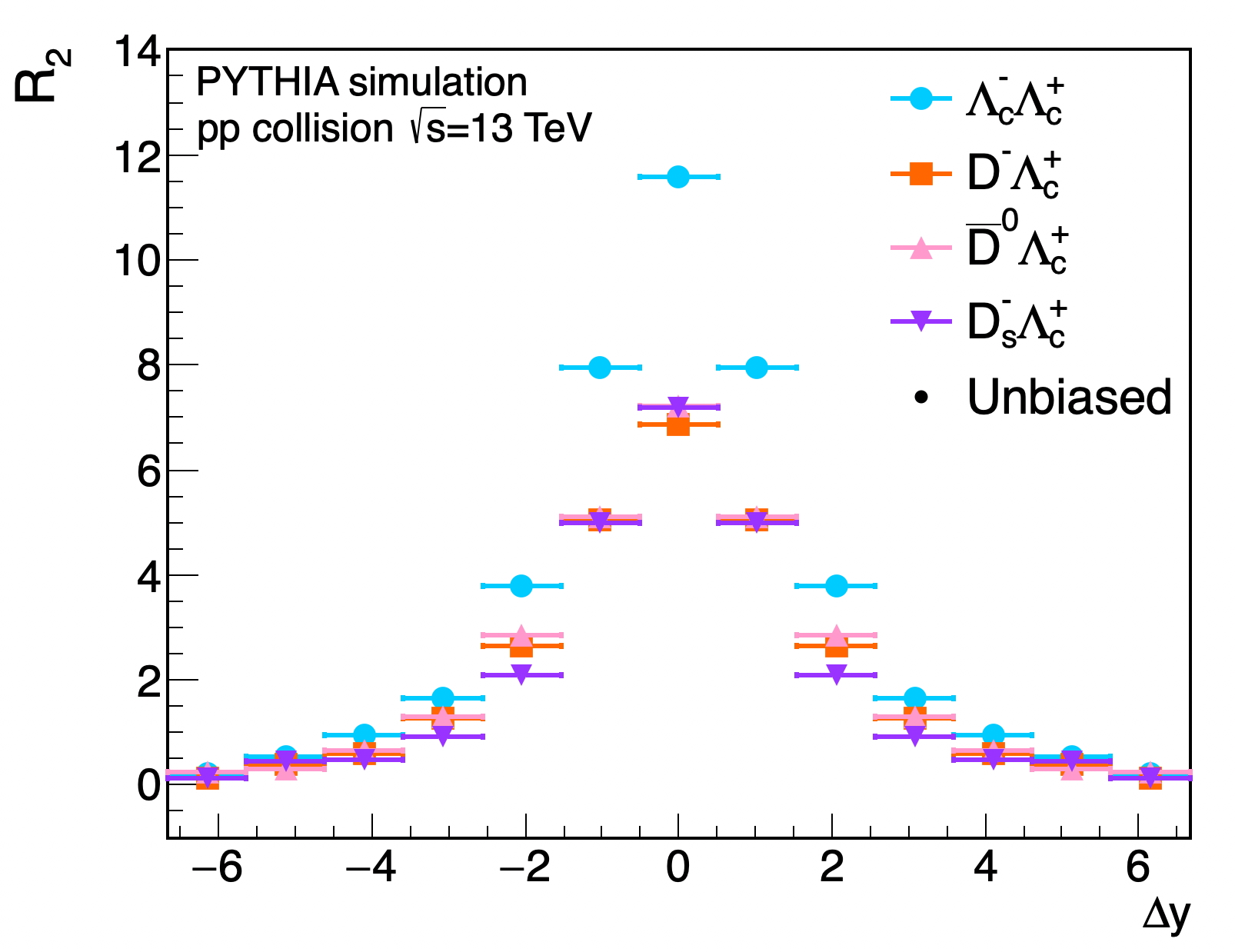}
 \caption{Projections of the normalized cumulant $R_2$ onto the $\Delta y$-axis for selected charm--charm hadron pairs in unbiased \textsc{Pythia} simulations of pp collisions at $\sqrt{s} = 13~\mathrm{TeV}$.}   
 \label{R2PorjectionUnbiased}
\end{figure}

Figures~\ref{A2PorjectionUnbiasedandBiased} and~\ref{R2PorjectionUnbiased} present projections of two-particle cumulants onto the rapidity difference axis $\Delta y$ for selected charm--charm hadron pairs. In both figures, $\mathrm{\Lambda_c^+}$ is used as the reference particle. Figure~\ref{A2PorjectionUnbiasedandBiased} displays the cumulant $A_2(\Delta y)$ (left panel) for unbiased events and the highly biased events, $n = 8$ (right panel). Since $A_2$ is normalized by the single-particle yield of the reference particle, it is sensitive to particle ordering, making it less suitable for direct cross-species comparisons. For example, $A_2(\Delta y)$ for the $\mathrm{D^0}$--$\mathrm{\Lambda_c^+}$ pair appears more pronounced than $\mathrm{\Lambda_c^-}$--$\mathrm{\Lambda_c^+}$ when $\mathrm{\Lambda_c^+}$ is the reference particle, owing to the relatively small yield of $\mathrm{\Lambda_c^+}$.

Complementing this, Fig.~\ref{R2PorjectionUnbiased} shows $R_2(\Delta y)$ for the same set of charm–charm pairs but in unbiased events, allowing for an assessment of baseline correlations. A flavor--antiflavor hierarchy is observed. Pairs like $\mathrm{\Lambda_c^-}$–$\mathrm{\Lambda_c^+}$, which share all three valence quark flavors ($c$, $u$, $d$), show the strongest correlation signal, indicative of $R_2$ sensitivity towards flavor balancing. Combinations such as $\mathrm{D^-}(d\bar{c})$--$\mathrm{\Lambda_c^+}(udc)$ or $\mathrm{\bar{D}^{0}}(u\bar{c})$--$\mathrm{\Lambda_c^+}(udc)$, which share two valence quarks ($c$ and $d$ or $c$ and $u$), exhibit a moderate peak. In contrast, $\mathrm{D_s^-}(s\bar{c})$--$\mathrm{\Lambda_c^+}(udc)$, which shares only the charm quark and includes a strange quark, shows a similar correlation magnitude to the $\mathrm{D^-}$ and $\mathrm{\bar{D}^{0}}$ cases. When comparing charm--charm pairs, the correlation strengths are remarkably similar, indicating that the charm quark content is the dominant factor shaping the $R_2$ structure. Ultimately, when comparing the magnitude of these correlations, they are larger than $\mathrm{\Lambda - \Lambda}$ correlations \cite{Patley:2024egb}, emphasizing the strong charm--anticharm correlation. While strange hadrons show a monotonic trend with the number of shared valence flavors, charm hadrons such as $\mathrm{D_s}$, $\mathrm{D^0}$, and $\mathrm{D^\pm}$ exhibit similar correlation strengths regardless of whether one or two valence quarks are shared. However, the baryon--baryon correlation $\mathrm{\Lambda_c^- - \Lambda_c^+}$ stands out with a significantly stronger signal. For light hadrons, a clear monotonic trend is observed \cite{Patley:2024egb}: correlations are strongest for three balanced flavors and decrease for two and one. In the charm sector, however, this pattern is modified since charm itself dominates the correlation structure, overpowering the contribution from the number of shared light flavors. While light quarks arise from non-perturbative Lund string breaking governed by hadronization parameters, charm quarks are produced early in hard scatterings, independent of such parameters. The comparison quantifies the interplay between light-quark contributions to flavor balancing and the significantly stronger intrinsic correlations of charm quarks.
\begin{figure}[htbp]
  \centering
  \vspace{2pt}
  \includegraphics[width=\columnwidth,clip]{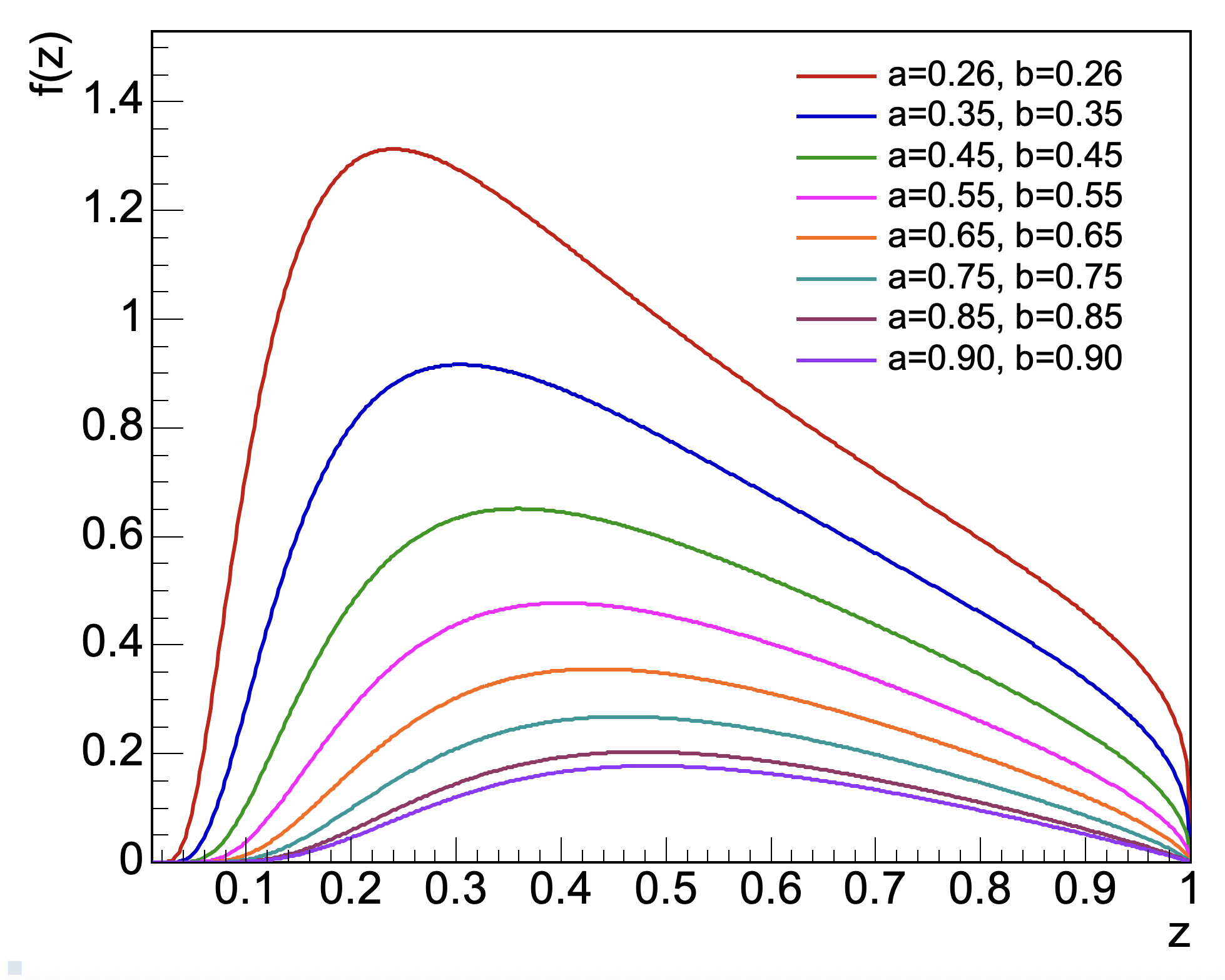}
  \caption{Variation of the fragmentation function 
        $f(z) \propto z^{-1}(1 - z)^a \exp\left( -b\, m_T^2/z \right)$ for selected values of the Lund parameters $a$ and $b$. The parameter $a$ controls the behavior as $z \to 1$, while $b$ introduces exponential suppression at small $z$. This function models how a charm quark hadronizes with light quarks during string breaking in the Lund model.
    }
  \label{lundfrag}
\end{figure}
\begin{figure*}[htbp]
  \centering
  \vspace{2pt}  
  \includegraphics[width=0.96\textwidth]{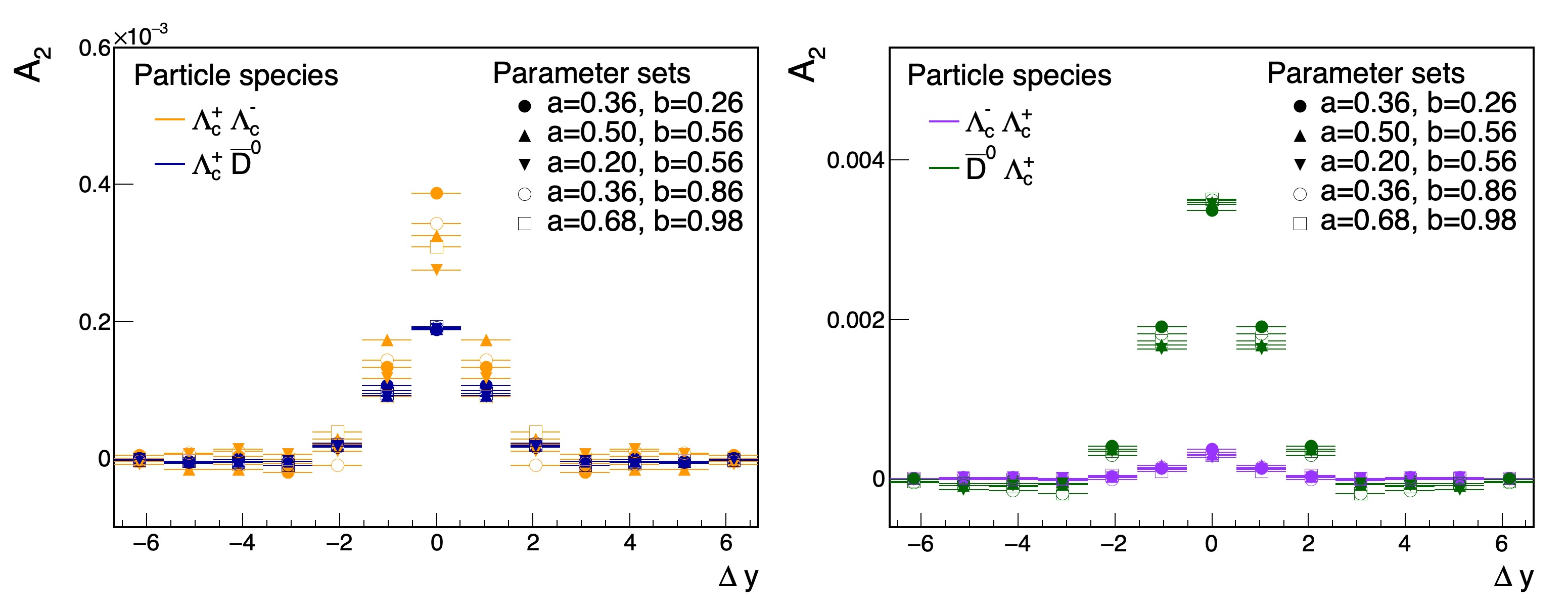}
\caption{Rapidity dependence of the correlation function $A_2(\Delta y)$ for various charm hadron pairs in biased events ($n=8$). Results are shown for different Lund fragmentation parameter sets to illustrate their impact on the correlation structure.}
  \label{lund}
\end{figure*}

How do $\mathrm{D^0}$ and $\mathrm{\Lambda_c^+}$ compete in production given their light-quark content? In the Lund string model, hadronization proceeds via string stretching and breaking, where the production of light \( u \) and \( {d} \) quarks are governed by model parameters. While charm quarks originate from early, high-energy hard scatterings and are unaffected by Lund tuning, the light-quark content that binds with charm is sensitive to hadronization settings. Therefore, comparing charm hadron species gives a handle on how light quarks are produced and binding with charm quarks. We investigated the impact of two parameters that control the fragmentation of Lund strings into hadrons. 

The first parameter, known as $a$ in the \textsc{Pythia} manual, regulates the hardness of the fragmentation, whereas the second parameter, $b$, suppresses the production of large transverse mass hadrons through an exponential damping factor. Figure~\ref{lundfrag} illustrates the dependence of the Lund fragmentation function on these two parameters. Although \textsc{Pythia} has been calibrated to measured data, the precise values of these parameters are not tightly constrained and may vary depending on tuning strategy and physics assumptions. It is thus of interest to find out whether correlations of hadrons, particularly charmed hadrons, could provide a probe to better tune these parameters. We have thus studied the sensitivity of the shape and strength of correlations to the magnitude of these parameters.
Selected results from our study are displayed in Fig.~\ref{lund}, which presents associated distributions $A_2$ obtained for $\mathrm{\Lambda_c^- - \Lambda_c^+}$ and $\mathrm{\bar{D}^0 - \Lambda_c^+}$ based on simulations performed with parameter values in the range $0.2 < a < 0.68$ and $0.26 < b < 0.98$. Table~\ref{tab:rms_lund} shows the RMS widths of $A_2$ for the pairs. The results indicate a notable sensitivity of the correlation widths to variations in the \textsc{Pythia} parameters $a$ and $b$ for the baryon--baryon configuration ($\approx 0.5$–$0.9$), and a small impact on the meson--baryon configuration ($\approx 0.89$–$0.95$). 

%
 \begin{table}[htbp]
  \centering
  \caption{RMS widths of $A_2$ distributions for $\mathrm{\Lambda_c^- - \Lambda_c^+}$ and $\mathrm{\bar{D}^0 - \Lambda_c^+}$ pairs at different \textsc{Pythia} parameter values.}
  \begin{tabular}{cc|cc|cc}
    \hline
    $a$ & $b$ & \multicolumn{2}{c|}{$\mathrm{\Lambda_c^- - \Lambda_c^+}$} & \multicolumn{2}{c}{$\mathrm{\bar{D}^0 - \Lambda_c^+}$ } \\
    \cline{3-6}
         &     & RMS & Error & RMS & Error \\
    \hline
    0.36 & 0.26 & 0.811 & 0.002 & 0.959 & 0.001 \\
    0.50 & 0.56 & 0.907 & 0.025 & 0.934 & 0.009 \\
    0.20 & 0.56 & 0.793 & 0.002 & 0.918 & 0.001 \\
    0.36 & 0.86 & 0.533 & 0.002 & 0.899 & 0.001 \\
    0.68 & 0.98 & 0.926 & 0.002 & 0.927 & 0.001 \\
    \hline
  \end{tabular}
  \label{tab:rms_lund}
\end{table}

We next consider the measurability of charm BF integrals. Charm is conserved by the strong interaction. The (strong) production of a charm hadron should thus always be balanced by the emission of an anticharm hadron. However, charm hadrons are known to decay on time scales ranging from $\sim 10^{-13}$ to $\sim 10^{-12}$ seconds. It is thus conceivable that observed charm-hadrons (or anti-hadrons) might not be manifestly balanced by the observation of anticharm (charm) hadrons because the balancing partner has decayed. 
The integral of the BF will then depend in part on the fraction of balancing partners that decay (hadronically or semileptonically) into non-charm hadrons. While this obviously impacts the charm BF sum-rule, it does so in a way that is model specific and depends on the relative production cross section of charm hadrons. A measurement of charm BF thus remains of interest. 

\begin{figure*}[htbp]
  \centering
  \vspace{2pt}  
  \includegraphics[width=0.96\textwidth]{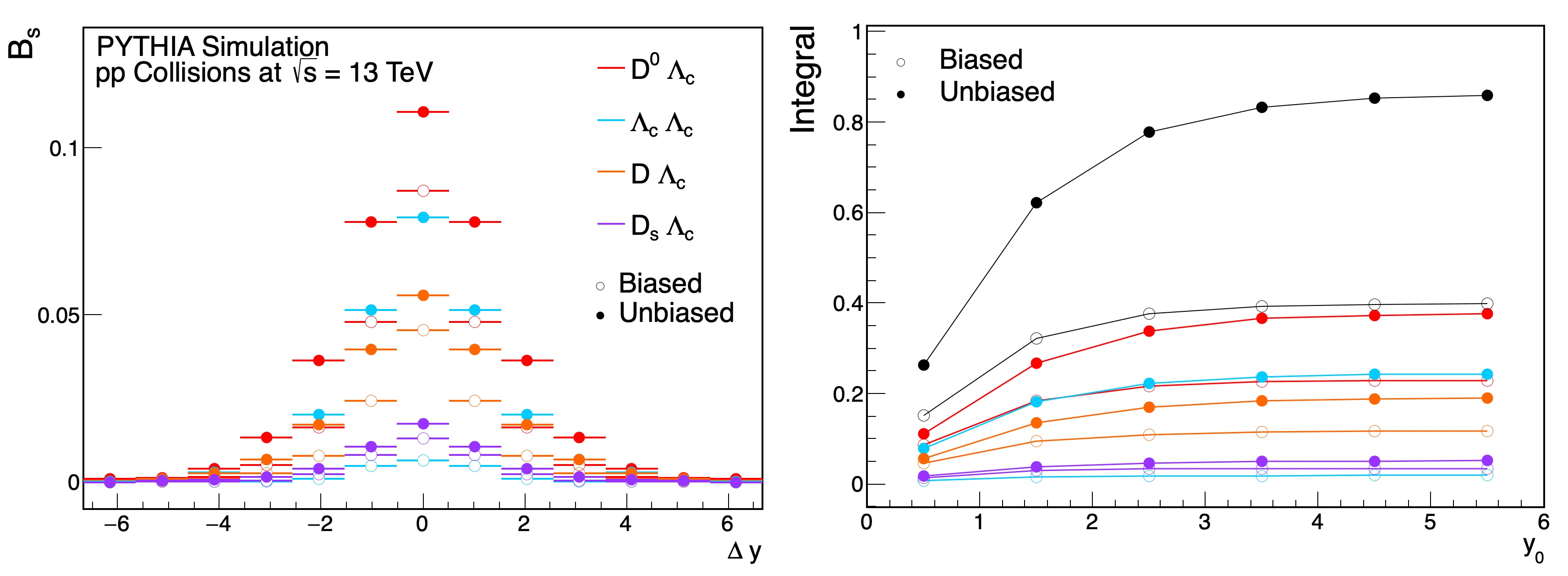}
\caption{
Balance function \( B_s(\Delta y) \) for charm--charm pairs with \(\mathrm{\Lambda_c}\) as the associated particle. 
Left: Balance function distributions as a function of the rapidity difference \(\Delta y\) for various trigger particles (\(\mathrm{D^0}\), \(\mathrm{D^+}\), \(\mathrm{D_s^+}\), and \(\mathrm{\Lambda_c^+}\)), shown separately for biased and unbiased \textsc{Pythia} events. 
Right: Cumulative integral of \( B_s(\Delta y) \) within a symmetric rapidity window \([-y_0, y_0]\), plotted versus \(y_0\), illustrating the dependence of the balance strength on the rapidity acceptance.
}
  \label{integral}
\end{figure*}

Comparison of charm BFs with theoretical predictions is evidently also impacted by the width of the rapidity acceptance used in a measurement. It is thus of interest to assess how the breadth of the acceptance impacts measurements of BF integrals. We address these two questions, in Fig.~\ref{integral}, based on plots of the BF vs. the longitudinal pair separation $\Delta y$ (left) and integral of the BF (right). BFs were computed for pairs involving a $\Lambda_c^+$ reference hadron. Associated hadrons considered include $\mathrm{D^0}$, $\mathrm{\Lambda_c^-, D^-}$, and $\mathrm{D_s}$. Although other higher charmed mass states exist, they would be challenging in the context of BF studies because the list of final states is open and inclusive. It should also be noted that the higher mass states typically decay into non-charm states. The balancing charm is thus indeed lost in these cases. The left panel displays symmetric BFs $B_s^{\alpha|\beta}(\Delta y)$ defined according to
\begin{equation}
\label{eq:BFs}
B_s^{\alpha|\beta}(\Delta y) = \frac{1}{2} \left[ B_2^{\alpha|\bar{\beta}}(\Delta y) + B_2^{\bar{\alpha}|\beta}(\Delta y) \right],
\end{equation}
whereas the right panel shows cumulative integral of $B_s$ expressed as a function of the acceptance width $y_0$. The BFs shown in Fig.~\ref{integral} were computed based on minimum bias events and biased events obtained with $n=8$. For illustrative purposes, no $p_{\rm T}$ thresholds and the full particle production  rapidity range were used for the computation of the BFs and their cumulative integrals. Experimental measurements would evidently feature practical limitations from finite $p_{\rm T}$ detection thresholds and pseudorapidity acceptance.  One notes that the BF with an associated $\mathrm{D^0}$ features the largest amplitude followed by amplitudes for $\mathrm{\Lambda_c^-}$, $\mathrm{D}$, and $\mathrm{D_s}$. The specific proportions of these BFs are best considered in the right panel of the figure, which shows the cumulative integral of the respective BFs of these four associated hadrons. Note how the BF integrals quickly rise for $y_0<2.5$ and then saturates beyond that. Taking these \textsc{Pythia} predictions at face value, this means that very good assessment of BF integrals could be obtained with acceptances of $-2.5 < y < 2.5$ which, not being available in ALICE 2 and difficult to achieve in CMS, would be ``easily" afforded in planned upgrades of these experiments.  One additionally notes that while  balancing the $\mathrm{\Lambda_c^+}$ with a $\mathrm{D^0}$ is the most probable (according to \textsc{Pythia}), it is closely followed by balancing with $\mathrm{\Lambda_c^-}$ and $\mathrm{D^-}$, whereas balancing with the production of  $\mathrm{D_s}$ is much less probable, owing most likely to the need to pick up a strange quark in the process. We also note that the actual balancing integrals (for very wide acceptance) are likely model specific and thus constitute, in their own right, good probes of the physics ingredients and their  implementations in specific models. Finally, also note that the sum of the BFs cumulative curves, shown as black dots in Fig.~\ref{integral}, saturate at a value of approximately 85\%. As anticipated, the sum of BF integrals does not add to unity. This is not a flaw of concept. Charm must be balanced by the strong interaction. However, weak decays introduce a charm  ``leak" or lack of balance. Higher mass charm states, not included in the calculations, that would fulfill the sum rule decay weakly on very short time scales. It would thus be extremely challenging to detect them and include them in measurement of charm BF integral. While this presents an additional experimental challenge,  it is clear that the fraction of balancing observable partners and the dependence of the BF on the acceptance measurement could be studied in details experimentally thereby enabling detailed assessment of the performance of theoretical models.

Figure~\ref{integral} additionally shows that applying a momentum transfer bias reduces both the strength and the integral of the charm--anticharm balance function. The analysis is based on an inclusive list of reconstructed hadrons, so decays from higher excited states may contribute indirectly and the result depends on the composition of the hadron list. In the unbiased sample, the integral approaches about 85\%, suggesting that a large fraction of charm--anticharm pairs hadronize into the selected species —$\mathrm{D^0}$, $\mathrm{D^\pm}$, $\mathrm{D_s^\pm}$— when the reference particle is $\mathrm{\Lambda_c^{\pm}}$. In the biased sample, the integral saturates at a lower value, around 40$\%$, which may indicate that higher momentum transfer in the initial production leads to a greater population of excited or non-included charm states. This trend shows how models reflect the impact of momentum bias and charm leakage through the inclusive hadron list in the final state.

\section{Conclusions}
\label{sec:conclusion}

We presented phenomenological studies of charm hadron correlations and balance functions (BFs) in pp collisions at $\sqrt{s}=13$ TeV based on the \textsc{Pythia} model. 
The flavor balancing sharing impact on the strength of the charmed correlations was studied and compared to the results for strange hadrons~\cite{Patley:2024egb}. \textsc{Pythia} predicts that heavy quarks should dominate over the flavor balancing shown by light quarks. We also studied the role of light flavor generation as well as the impact of variations of Lund string fragmentation  parameters. We find these  notably affect the width of the correlations, particularly in the $\mathrm{ \Lambda_c^- -\Lambda_c^+}$ channel, while the effect is more modest for the $\mathrm{\bar{D}^0 - \Lambda_c^+}$ correlations. 
We thus conclude that  the charm correlations exhibit good sensitivity to the details of the string fragmentation model. Additionally, we find that  \textsc{Pythia}  predicts a strong charm balancing ranking. For instance, the emission of a $\mathrm{\Lambda_c^+}$ is expected to be balanced  by a $\mathrm{D^0}$ with a probability of $\sim 45$\%, whereas charm balancing emissions by $\mathrm{\Lambda_c^-}$, $\mathrm{D^-}$, and $\mathrm{D_s}$ contribute approximately $30\%$, $20\%$, and $5\%$, respectively.  \textsc{Pythia}  additionally predicts that in the absence of a momentum-transfer bias, approximately 85\% of the $\mathrm{\Lambda_c^+}$ particles are balanced by the inclusive set ${\mathrm{\Lambda_c^-},\ \mathrm{D^0},\ \mathrm{D^\pm},\ \mathrm{D_s}}$. This fraction goes down appreciably for high momentum transfer collisions suggesting a complex dependence of the charm balancing on the $p_{\rm T}$ scale of the interactions. 

Future measurements of charm correlation functions and balance functions proposed  in this work shall shed new light on charm production and hadronization. The feasibility of such measurements would  depend on the acceptance and performance of experiments as well as the size of the acquired data sets. In this work,
simulations were carried out for an ideal experimental setup with wide pseudo-rapidity acceptance, no transverse momentum threshold, and no efficiency losses. Current experiments are however limited based on their pseudorapidity and transverse momentum acceptance, their displaced vertex resolution, as well as overall charm hadron reconstruction efficiency. Generator-level studies using 50 billion events suggest that achieving similar statistical precision with reconstructed data would require datasets far exceeding 50 billion events, depending on charm reconstruction efficiency. 
Future collider experiments with wider pseudorapidity acceptances, enhanced particle identification, and drastically improved vertex resolution, such as the planned ALICE~3~\cite{ALICE:2022wwr,Nicassio:2023zoj} and CMS upgrade~\cite{FernandezPerezTomei:2024xid}, will make precision charm balance function measurements significantly more practical.

\section*{Acknowledgments}
The authors wish to dedicate this work to the memory of our colleague and friend Dr.~Sumit~Basu. This work was supported in part by grant No. DE-FG02-92ER40713, grant No. SR/MF/PS-02/2021-IITB  (E-37126), and grant No. PN-III-P4-PCE-2021-0390.
\newpage

\bibliography{reference.bib}

\end{document}